\documentclass[runningheads]{llncs}

\usepackage{amsmath,amsfonts,bm}
\usepackage[dvipsnames]{xcolor}
\usepackage{tikz}

\definecolor{MyRed}{rgb}{0.6,0.0,0.0} 
\definecolor{MyBlack}{rgb}{0.1,0.1,0.1}

\newcommand{\sstitle}[1]{\smallskip\noindent\textbf{#1.\/}}

\newcommand{\sititle}[1]{{\it #1:\/}}

\def\eqref#1{equation~\ref{#1}}
\def\1{\bm{1}}

\def\vc{{\bm{c}}}

\def\ve{{\bm{e}}}

\def\vs{{\bm{s}}}
\def\vt{{\bm{t}}}

\def\vz{{\bm{z}}}

\def\mM{{\bm{M}}}
\def\mN{{\bm{N}}}

\def\mW{{\bm{W}}}

\DeclareMathAlphabet{\mathsfit}{\encodingdefault}{\sfdefault}{m}{sl}
\SetMathAlphabet{\mathsfit}{bold}{\encodingdefault}{\sfdefault}{bx}{n}

\def\sC{{\mathbb{C}}}

\def\sR{{\mathbb{R}}}

\def\sT{{\mathbb{T}}}

\def\emM{{M}}

\newcommand{\Ls}{\mathcal{L}}

\usepackage{booktabs}
\usepackage{graphicx}
\usepackage{hyperref}
\usepackage[T1]{fontenc}
\usepackage{mathtools}
\makeatletter \newif\if@restonecol \makeatother  
\usepackage[linesnumbered, ruled, vlined]{algorithm2e}
\usepackage{algpseudocode}

\usepackage[bottom]{footmisc}
\usepackage{epstopdf}
\usepackage{graphics}
\DeclareGraphicsExtensions{.eps,.gz,.eps,.pdf,.png,.jpg}
\usepackage{makecell}
\usepackage{listings}
\usepackage{graphicx}
\usepackage{amsmath,amssymb,amsfonts}
\usepackage{dsfont}
\usepackage{wrapfig}
\usepackage{enumerate}
\usepackage{paralist}
\usepackage{multirow}
\usepackage{verbatim}
\usepackage{lipsum}
\usepackage{url}
\usepackage{times}
\usepackage{mathrsfs}
\usepackage{threeparttable}
\usepackage{tabularx}
\usepackage[dvipsnames]{xcolor}
\usepackage{capt-of}

\hypersetup{
  bookmarks=true,        % show bookmarks bar?
  unicode=true,          % non-latin characters in bookmarks?
  pdffitwindow=false,    % window fit to page when opened
  pdfstartview={FitH},   % fits width of the page to the window
  pdftoolbar=true,       % show acrobat toolbar?
  pdfmenubar=true,       % show acrobat menu?
  colorlinks=true,      % false: boxed links; true: color links
  linkcolor={blue!80!black},       % color of internal links
  urlcolor={blue!30!black},        % color of external links
  citecolor={green!60!black},       % color of links to bibliography
  filecolor=black,       % color of file links
  pdftitle={},           % title
  pdfauthor={}, % author
  pdfcreator={} % creator of pdf
}

\begin{document}
\title{From Scattered Sources to Comprehensive Technology Landscape
        :\\ A Recommendation-based Retrieval Approach
        }
%
%From heterogenous sources to comprehensive technology landscape: an information retrieval approach

%author's information
\author{\textsc{Duong} Chi Thang\inst{1} \and
\textsc{Percia David} Dimitri\inst{2;3} \and
\textsc{Dolamic} Ljiljana\inst{3} \and
\textsc{Mermoud} Alain\inst{3} \and
\textsc{Lenders} Vincent\inst{3} \and
\textsc{Aberer} Karl\inst{1}}
%%
%\authorrunning{Duong et al.}
\institute{Distributed Information Systems Laboratory, EPFL \and Information Science Institute, University of Geneva \and Cyber-Defence Campus, armasuisse Science and Technology, EPFL}
\maketitle              % typeset the header of the contribution
\begin{abstract}
Mapping the technology landscape is crucial for market actors to take informed investment decisions. However, given the large amount of data on the Web and its subsequent information overload, manually retrieving information is a seemingly ineffective and incomplete approach. In this work, we propose an end-to-end recommendation based retrieval approach to support automatic retrieval of technologies and their associated companies from raw Web data. This is a two-task setup involving (i) technology classification of entities extracted from company corpus, and (ii) technology and company retrieval based on classified technologies. Our proposed framework approaches the first task by leveraging DistilBERT which is a state-of-the-art language model. For the retrieval task, we introduce a recommendation-based retrieval technique to simultaneously support retrieving related companies, technologies related to a specific company and companies relevant to a technology. To evaluate these tasks, we also construct a data set that includes company documents and entities extracted from these documents together with company categories and technology labels. Experiments show that our approach is able to return 4 times more relevant companies while outperforming traditional retrieval baseline in retrieving technologies.
% significantly better results in both technology and company retrieval

\end{abstract}
\section{Introduction}
%\thang{I remove the author list since WISE is double blind}
%\dola{Maybe: SwissTech}

The expanding and accelerating pace of technology development continuously reshapes the technological landscape \cite{shalf_future_2020}. Depicting an up-to-date and holistic map of organizations that develop, implement or sell a given technology is an important business challenge, should a technology be novel or established \cite{rikhardsson2018business}. In such a dynamic environment, market actors are increasingly confronted by an information overload issue, as an ever raising amount of heterogeneous and unstructured market information needs to be collected, stored, cleaned, structured and analyzed \cite{roetzel2019information}. Hence, the automated analysis of the complex network of organizations and technologies is a key business intelligence necessity not only for public entities, but also for private investors~\cite{durak2018flight}.

Such an information retrieval necessity has triggered various business intelligence and technology monitoring procedures, which have been either developed by in-house R\&D efforts of organizations or by academic actors (\textit{e.g.}, \cite{daim2016anticipating}). Often, such procedures consist of non-automated approaches that struggle to tackle the information overload challenge, as they do not provide a reliable, systematic and scalable information retrieval methodology for mapping the technology market and determine which companies are developing/commercializing which technology \cite{loveridge2016fta}. These extant procedures are mainly based on frameworks that find documents matching query terms instead of finding entities \textit{per se} \cite{demartini2009vector}. Yet, finding entities is central when it comes to investigate the technological landscape and finding new technologies. Even though related works have partially investigated such an issue, to the best of our knowledge, no work has provided an end-to-end automated information retrieval framework for finding technology related entities and mapping the technological landscape through a recommendation based perspective. In this work, we develop such a framework in order to map the industrial technology landscape, which would be highly applicable in several domains where a comprehensive understanding of the landscape is required. For instance, in cybersecurity, by mapping the technological landscape of cybersecurity, decision-makers  will be provided relevant information for taking more informed purchase decisions\cite{mahmood2013security}. Similarly, in the stock market, having a comparative analysis on technological advances of competing companies is crucial for making correct investment decisions.

Our framework is a two step approach that first classifies the entities into technologies before performing company and technology retrieval. We use DistilBERT, which is a state-of-the-art language model, to construct the entity embeddings allowing us to achieve high accuracy in technology classification. For retrieval, we propose a recommendation based approach that takes into account both the technologies, companies and their relationship. This recommendation approach enables better retrieval results {in comparison with state-of-the-art learning based recommendation approaches such as GMF~\cite{he2017neural,rendle2009bpr}, MLP~\cite{he2017neural} and NCF~\cite{he2017neural}.}
%as we can answer queries by looking at similar companies or technologies. 
%In addition, it is also robust to missing data. 
Our results show that our approach is able to return 4 times more relevant companies in company to company retrieval in comparison with traditional tf-idf retrieval approach.

%employs... (to be completed by \thang{-> employed methodologies and steps [in a nutshell]}). We use an extant business-intelligence platform that employs a \textit{tf-idf} approach as a baseline, and then we compare the results of our framework with such a baseline. Our results show that our approach is able to retrieve significantly better results in both technology and company retrieval.\thang{to be completed}.

The remainder of this article proceeds as follows. Section \ref{sec:related} grounds this research by emphasizing different methodologies of information retrieval for technology monitoring, and the research gaps. Section \ref{sec:dataset} details the data collection. Section \ref{sec:model} presents the information retrieval model and framework. Section \ref{sec:tech_class} presents the entity extraction methodology and the technology classification we use. Section \ref{sec:retrieval} presents the technology retrieval method we used. Section \ref{sec:eval} presents the empirical evaluation, while Section \ref{sec:conclusion} concludes and set the path for further research.

%\thang{We need to answer 3 important questions: 1) Is there any existing technology retrieval approach? If not, why not? 2) What makes the technology retrieval problem different from traditional retrieval? 3) What are the use cases? What the users expect from a tech retrieval system? }

\label{sec:intro}

\section{Related Works}

The information overload triggered by the big data era has motivated researchers and practitioners to develop numerous automated information retrieval methods by using different yet often complementary approaches~ \cite{munir2018use,kaur2021comparative,shalaby2019patent}. Such methods have been widely used in fields as digital libraries~\cite{hersh2014information}, information filtering and recommender systems~\cite{belkin1992information,heggo2021data}, media search \cite{rao2019multi} and search engines~\cite{aggarwal2018information}. 

In the field of technology monitoring and forecasting -- and more specifically in the specific context of technology landscape monitoring --, numerous works have been published, involving the extraction of patents~\cite{kim2016generating}, scientific articles~\cite{dotsika2017identifying} and social media analysis~\cite{chen2019disruptive}. These technology retrieval methods can be classified into either a \emph{keyword-based} or a \emph{entity-based} approach. Most of the existing methods are keyword-based in which the queries and the data are mostly plain texts~\cite{Arai2020}.

Hossari et al. (2019) proposed an automated framework for detecting the existence of new technologies in texts, and extract terms used to describe new technologies \cite{hossari2019test}. Aharonson \& Schilling developed a framework that captures the distance between patents, and a company's technological footprint. Their framework also enables to measure the proximity of technological footprints between organizations \cite{aharonson2016mapping}. Tang and Liu (2008) presented a three-layer model for technology forecasting based on text mining techniques, incorporating a collection layer, an analysis layer and a representation layer, before overlapping a semantic web based approach in order to map the industrial technology landscape \cite{tang2008applying}. On the other hand, entity-based techniques \cite{wang2017leveraging} require linking a piece of text to an entity in a knowledge base such that retrieval is done on the entities instead of the raw texts. Woon and Madnick (2008) developed an information retrieval framework for visualizing the technology landscape by exploring the use of term co-occurence frequencies as an indicator of semantic closeness between pairs of entities \cite{madnick2008semantic}. In the field of energy related technologies, Mikheev (2018) developed an ontology based data access framework under a semantic approach to query complex datasets, creating an automated mapping procedure to connect data to ontology entities \cite{mikheev2018ontology}. By applying a semantic approach, Sitarz et al. (2012) developed an automated framework for identifying thematic groups of scientific publications based on clustering of sets of co-occurence words and and financial-analysis techniques for trends detection and forecasting~\cite{sitarz2012application}. 

In our setting, we opt for the entity-based approach as it allows us to handle different mentions of the same entity while enabling better retrieval accuracy due to external information from the knowledge base. However, to the best of our knowledge, little has been done when it comes to apply information retrieval methodologies with the aim of presenting a holistic and comprehensive monitoring and mapping of the industrial technology landscape. In order to do so, a technique needs to consider both the technologies and companies at the same time. The following steps have to be undertaken: (i) an entity fishing approach needs to be applied for extracting and classifying technology entities; (ii) then, these technology entities need to be linked to specific companies; (iii) and finally, these entities must be ranked according to their level of relevance. Demartini et al. (2009) provided a formal model for entity fishing and ranking \cite{demartini2009vector}. Yet, to the best of our knowledge, no such work has been deployed in the context of technology landscape monitoring. Similarly, Balog et al. (2012) developed a framework for assessing the strength of association between a topic and a person ( \textit{i.e.}, expertise retrieval) \cite{10.1561/1500000024}. Yet, to the best of our knowledge, no such framework has been applied in the context of technology landscape monitoring for linking technology entities and company entities.

\label{sec:related}

\section{Dataset construction}
\label{sec:dataset}

{To the best of our knowledge, there is no public dataset available for the technology retrieval and classification task. As a result, we aim to create and publish such a dataset to further research in this field. We have defined the following requirements for our dataset. First, the dataset should be multilingual as the technology retrieval task should be language independent. Second, the dataset should be realistic and coming from real-world data as this would enable objective evaluation of any proposed approach. }
%Third, the dataset construction should be done in an automatic manner as it enables reproducibility and possible future extension. }

{In the following, we discuss our data construction process. The dataset is constructed by first crawling different data sources that are publicly available on the Internet.}
As we aim to develop a language-agnostic framework, we decide to construct a dataset based on Swiss companies as Switzerland is a multilingual country with French, German and Italian as official languages while English is a working language. For each company, we intend to collect all possible documents that are related to the company's actual activities. This involves the following data sources: the company's \emph{website}, the company's \emph{job postings}, the \emph{patents} and the company's \emph{tweets}. In addition, to maintain an up-to-date dataset about the companies, we periodically crawl data from the above data sources. {The above data sources are selected as they could provide different perspectives regarding a company's technology offering. }
{The statistics of the dataset\footnote{Note that there are overlapping companies and terms among different data sources.} is shown in \autoref{tbl:stats}.}

\begin{table}[t]
\centering

\caption{Dataset statistics.\label{tbl:stats}}
\begin{tabular}{lrr}
\hline
\textbf{Dataset} & \multicolumn{1}{l}{\textbf{\#comps.}} & \multicolumn{1}{l}{\textbf{\#terms}} \\ \hline
Website & 7907                            & 22104                       \\
Patent  & 6814                            & 7796                        \\
Jobs  & 3894                            & 2532                        \\ 
Twitter & 790                             & 3236                        \\ \midrule
\textbf{Total}   & 18339                           & 27977                       \\ \hline
\end{tabular}
\end{table}

\subsection{Collecting data}
To collect the data, we first need a list of Swiss companies. We obtain the list of companies registered in Switzerland from the federal Central Business Name Index (Zefix)\footnote{https://zefix.ch}. Additional information regarding given company is then extracted from the corresponding cantonal commercial register record. These records provide, among others, information on location, people, type of company.  

\sstitle{Websites}
As commercial registers do not provide the information on regarding websites of the registered companies, we need to find the company's website based on the company's name. We first clean the company name (e.g. removing Ltd.) before performing Google search. The information extracted from commercial registers in combination with the results of the previously mentioned search are then put through the classifier to link the company to a correct website. We then crawl the pages from the detected website, using their text for entity extraction. 
%The information regarding the company's contact information and social media accounts is also extracted from the website content.      

\sstitle{Patents}
We use Patents data from United States Patent and Trademarks Office (USPTO). PatentView\footnote{https://patentsview.org} provides half yearly data dump of the USPTO database, which we inject directly into our system. The patents are linked to the companies based on the location and assignee information, while the title and the abstract of the patents are used to extract the entities. 

\sstitle{Jobs}
The Indeed\footnote{http://indeed.ch} is used as a source of information related to jobs. We perform weekly per canton search for jobs, retrieving for each job the following information: title, description, company name and the original posting. Job's title and description are used for technology annotation, while the linking to the company is based on it's name and location.   

\sstitle{Tweets}
For each company, we look up its Twitter handle from Crunchbase which is a commercial database of company information. From the handle, we collect the latest 3000 tweets for each company using sempi.tech which is a social listening framework. We use these tweets for entity extraction.

\subsection{Entity Extraction}
%In this section, we describe our method of extracting named entities from different data sources. These entities are potential candidates for our technology classification step. 
From the documents collected in the previous step, we use DBPedia Spotlight (DBPS)\cite{10.1145/2506182.2506198}, an open source tool to automatically annotate the mentions of the DBPedia entities within the text content. 
Dbpedia is a multi-domain ontology derived from Wikipedia. 
Each DBPedia entity is an URI with the prefix \textit{http://dbpedia.org/resource/} followed by the identifier of the corresponding Wikipedia article.
DBPS allows us to not only extract entities but also link the entities to their corresponding DBPedia entities. DBPS is also capable of handling multiple languages, which is important in our setting.  
For each data source, we perform entity extraction independently. The output of this step is a list of entities and their corresponding number of appearance in the data source of a company. We store the output in JSON line format where each line is a triple of company, DBPedia entity and number of entity occurrences. {Note that while our data sources are heterogeneous as they come from different domains, DBPS allows us to standardize them. Whether it is a website, a patent, a job posting or a tweet, DBPS extracts all the candidate technology mentions while ignoring irrelevant data. These technology mentions are the input to any technology classification or retrieval technique.}

%We use DBPedia Spotlight~\cite{mendes2011dbpedia}, an open source tool to automatically annotate the mentions of the DBPedia entities within the text content of the company's website.

\subsection{Data labeling}

\sstitle{Technology labeling} To support the evaluation of a technology classifier, we also provide a list of Wikipedia terms and their labels whether a term is about a technology or not. 
Even though Technology is one of Wikipedia's Main Topic Classification (MTC) categories, one can not rely on this concept to extract all technology related articles, as the categories within the Wikipedia graph are very loosely related("is related to"). We approach the labeling in a top-down approach where we aim to label the MTC categories. First, Wikipedia directed categories graph was cleaned by removing hidden categories, admin and user pages, followed by regular expression filters removing categories referring to companies, brands, currencies etc. We than calculate the shortest path to each of the 28 MTC, and retain the categories having the shortest path to Technology, Science, or Engineering topics. This process resulted in the list of 7876 categories. These categories were than manually labeled as technology or non technology, resulting with 1356 categories being labeled an technology. 
{This is the only manual step in our data construction process.}
An article is then considered to be a technology if it is directly connected to a category labeled as such. 

%Note that Wikipedia articles are organized into \emph{categories} and \emph{entities}. We first go over all English Wikipedia categories and manually select \todo{...} Wikipedia categories that are technological. We then label all the Wikipedia entities that belong to these categories as technologies. This results in \emph{...} technologies. 

\sstitle{Company categorization} Crunchbase maintains a database of companies and their detail information where the data are manually curated by Crunchbase staff and online contributors~\footnote{crunchbase.com}. Each company is associated with several categories describing its main activities. For instance, Roche which is a pharmaceutical company based in Switzerland is categorized as Biotechnology, Health Care, Health Diagnostics and Pharmaceutical. From the companies collected in the above steps, we crawl Crunchbase to obtain their categories. The categories can be considered as pseudo-labels for our com-com and tech-com retrieval tasks.

\section{Model and Approach}
\label{sec:model}
We develop a unified framework to classify entities into technologies and to perform technology related retrieval. This requires solving two tasks of technology classification and technology retrieval.

\subsection{Model}
Our framework considers a set of companies $C = \{c_1,\cdots,c_n\}$ and a set of entities $E = \{e_1,\cdots,e_m\}$. The connection between an entity $e_i$ and a company $c_j$ can be observed through several data sources. For each data source, we measure this connection by the number of times the entity $e_i$ is mentioned by the company $c_j$. We can present these connections for a specific data source by an \emph{interaction matrix} $\mM$ where $\emM_{ij}$ is the occurrence frequency of the entity $e_i$ in the corpus of company $c_j$. We also denote these matrices by $M=\{\mM_1,\cdots,\mM_k\}$ where $\mM_i$ represents an interaction matrix. 

\sstitle{Tasks} We have 3 retrieval tasks: company to technology (com-tech) retrieval, company to company (com-com) retrieval and technology to company (tech-com) retrieval. Com-tech and com-com retrieval are connected as both require an accurate representation of a company by its technologies. Two companies are similar if they have similar technology representation. 
Similarly, for tech-com retrieval, we need a good representation of a technology by its companies. There is a mutual reinforcing relationship between company and technology representation. A good company representation requires knowing similar technologies while knowing company similarity is helpful in constructing a good technology representation. 
This means we need a common model to approach these three different retrieval tasks.
%While there are 3 different retrieval tasks in our setting, the com-tech retrieval is the most important one as it provides the foundation for the other tasks. A good com-tech retrieval corresponds to a good representation of a company by its technologies, which is crucial in finding similar companies for the com-com task. On the other hand, tech-com searches can be done by reversing the results from com-tech retrieval. 
%Technology-company retrieval is inherently the reverse retrieval of company-technology retrieval while having good results for company-technology retrieval 
%we focus on the company-technology retrieval task as solving this task makes the other retrieval tasks straightforward. 

\subsection{General approach} %\name~works in several steps as illustrated in \autoref{fig:architecture}. In the first step, we extract the entities from the company corpuses using two entity extraction frameworks, which can handle different languages. These frameworks also take care of entity disambiguation and entity linking with DBPedia and Wikidata. The results of the first step are a list of entities with their detail information. 

Our framework takes as input the entities extracted from the company corpora. These entities are identified using entity extraction frameworks such as DBPedia-spotlight. Each entity is associated with its description. 
This information is used in the second step to classify the entities into technologies or not. Our technology classifiers are constructed using BERT as a feature extractor. 
%BERT, which is a state-of-the-art language model, allows us to capture 
The technologies obtained from the previous step are used as input for the retrieval step. We reformulate the retrieval problem as a recommendation problem based on collaborative filtering where the technologies are ``recommended'' to a company if the technologies are considered to be related to the company's activities. 
Casting this as a recommendation problem has several benefits.  The relevancy of a technology to a company can be measured more accurately if similar companies in the same domain are considered. This also means that missing technologies in a company corpus can be recovered by considering similar companies. We propose a technology recommendation model that extends traditional matrix factorization by integrating both the technology meaning and the interactions between technologies and companies. 

% 1) it enables us to handle missing data in the corpuses, 2) the retrieval results is better as we can answer queries based on company and technology similarity. 

\begin{figure}
    \centering
    \vspace{-0.5em}
    \includegraphics[width=1.0\linewidth]{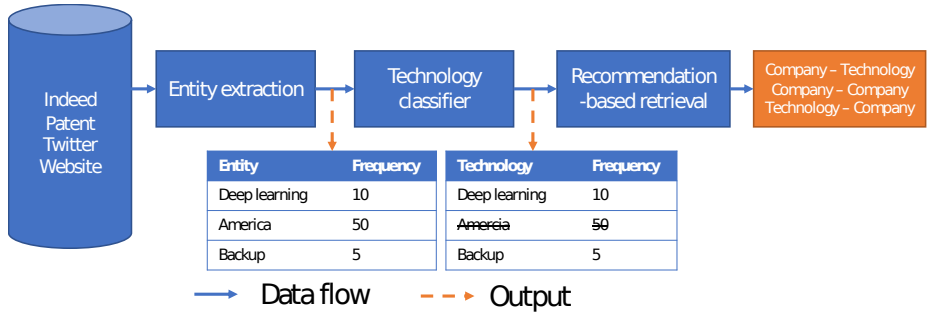}
    \caption{From data sources to company and technology retrieval}
    \label{fig:architecture}
    \vspace{-2.5em}
\end{figure}

\section{Technology Classification}
As the input entities to our system are extracted using an entity extraction framework such as DBPedia-spotlight~\cite{mendes2011dbpedia}, they cover all possible domains while we are interested only in technologies. To this end, we develop a technology classifier to filer out unrelated entities. 
Note that DBPedia-spotlight also performs entity linking where each entity is linked to a DBPedia page or a Wikipedia article describing this entity. 
We leverage this description to construct our technology classifier. 
For each entity, we extract its abstract which we define to be the description from DBPedia or the first paragraph of its Wikipedia article whichever available.

\sstitle{BERT-based encoder} We use DistilBERT~\cite{sanh2019distilbert} which is a state-of-the-art language model while being light-weight and fast to train to construct abstract embeddings. We use these abstract embeddings as representations for the entities. We pass each abstract through DistilBERT to obtain the token embedding of the [CLS] token, which is commonly used as the sentence or paragraph embedding. We denote this token embedding as $\vz_e$ where $e$ denotes an entity. 

Although DistilBERT is a distilled version of BERT which was trained on English wikipedia, distillation may incur information lost. To this end, we propose to fine-tune DistilBERT to better capture the abstract meaning which usually contain several scientific terminology, therein, obtaining better entity representation. 
{We fine-tune DistilBERT for our technology classification in an end-to-end manner where we feed the abstracts through the model to obtain the abstract embeddings. These abstract embeddings are then fed through a linear classifier to get the technology predictions. The parameters of DistilBERT and the linear classifier are updated together in an end-to-end manner using SGD~\cite{ruder2016overview} with the binary cross entropy loss as the loss function.}
We observe that with finetuning, we are able to achieve better accuracy.

\sstitle{Embedding refinement} To further increase the capacity of our classifier, we pass $\vz_e$ through several layers including a dropout layer to reduce overfitting. More precisely, $\vz_e$ is passed through a multi-layer neural network with 2 blocks where each block is a linear layer followed by a BatchNorm layer and a sigmoid non-linearity. A block can be formulated as follows:
\[\vt_e =\sigma(BatchNorm(\mW\vz_e + b_1))\]
Between 2 blocks, we also use a dropout layer to reduce overfitting. We observe that be refining the embedding this way, we can achieve better accuracy than non-refinement. 
\label{sec:tech_class}

% \subsection{Putting it together}

\section{Technology Retrieval}
% In this section, we discuss our method of find relevant technologies and companies.

%\subsection{Recommendation-based retrieval}

\sstitle{The case for retrieving as recommending}  There are several requirements for the retrieval model of this task. First, the technologies retrieved for a company must be derived from the technologies mentioned in the company's corpus as these technologies are the most likely ones to reflect the company's actual activities. As each technology has a different level of relevancy to the company, com-tech retrieval considering only observed technologies is a \emph{technology ranking} problem. 
Second, it is safe to assume that a company's corpus may not contain all the technologies the company is working on. Companies may not publicly mention a technology to keep a competitive advantage or it could simply be due missing data. For these reasons, the com-tech retrieval is also a \emph{technology discovery} problem. 
%We can infer potentially-related technologies for a company by looking at similar companies in the same domain. 
%By considering the activities of similar companies, we can identify potential related technology for a company of interest. On the other hand, the similarity between companies are measured based on their technology profile. 
To solve both problems at the same time, we propose a recommendation model that identifies potentially related technologies of a company by looking at similar companies while measuring the relevancy between every pair of technology and company for ranking technologies. 

\sstitle{Recommendation model} Given a set of companies $\sC =\{c_1,\cdots,c_n\}$ and the set of technologies $\sT = \{t_1,\cdots,t_m\}$ obtained after the technology classification step, our recommendation model takes as input a com-tech interaction matrix $\mM\in \sR^{n\times m}$ as 
\[
\emM_{ij} = \begin{cases}
	f(c_i,t_j), \text{if there is a mention of $t_j$ in any data source of $\sC_i$}\\
	\emptyset, \text{if there is no mention of $t_j$ in $\sC_i$}
\end{cases}
\]
The function $f(c_i, t_j)$ captures the importance of $t_j$ according to $c_i$. This importance can be measured by the number of times $t_j$ occurs in the corpus of $c_i$ rescaled by some weighting scheme. In our setting, we measure the importance per data source using tf-idf before combining all data sources:
\[
	f(c_i, t_j) = \sum_{k} w_k f_k(c_i, t_j)
\]
where $f_k(c_i, t_j)$ is the importance function of data source $k$ which is measured by the tf-idf value of $t_j$ considering each company as a ``document''. $w_k$ is its associated weight which captures our perceived relevance of the data source to our retrieval tasks.

A high value of $f(c_i,t_j)$ does not mean the company $c_i$ is actually working on technology $t_j$. For instance, during the pandemic, there could be several mentions of the word ``vaccine'' but it does not necessarily mean that a certain company is developing a vaccine. This example shows that answering com-tech retrieval by only looking at the company corpus could be problematic. 
We can have the same argument for $\emM_{ij} = \emptyset$. This does not mean the company $c_i$ has no activity related to $t_j$. It could be the case that the company is working on this technology but it has not mentioned it \emph{yet} in the corpus. 

The above com-tech interaction model is akin to collaborative filtering with implicit feedback~\cite{rendle2009bpr,he2017neural}. 
%While observed entries of $\mM$ captures the companies activities with related to technologies, the unobserved entries could just be missing data. 
To answer the com-tech retrieval problem, we first need to solve the recommendation problem where we need to estimate the unobserved entries of $\mM$. The estimation is usually done by learning a model $\hat{\emM}_{i,j} = f(c_i,t_j|\Theta)$ where $\hat{\emM}_{i,j}$ is the estimated score of $\emM_{ij}$, $f$ is a parameterized function that predicts the interaction score between $c_i$ and $t_j$, $\Theta$ denotes the parameters of $f$.

\sstitle{Semantic-aware matrix factorization} We propose a semantic-aware recommendation model extending traditional matrix factorization (MF) approach. In MF, each company and each technology is represented by an embedding in a shared latent space. The technology and the company embeddings are learned such that they can reconstruct the interaction matrix $\mM$. More precisely, let $\vc_i \in \sR^d$ and $\ve_j \in \sR^d$ be the embeddings of company $c_i$ and technology $t_j$ respectively. Then, MF aims to estimate the relevancy score $\emM_{ij}$ by:
\[\hat{\emM}_{ij} = f(c_i,t_j|\vc_i, \ve_j) = \vc_i^T\ve_j \]
%There are two reasons traditional MF is not suitable in our setting. First, the model capacity is controled by the embedding size $d$. Simply increasing $d$ is not 
However, MF considers the technologies to be independent even if they are semantically related such as ``deep learning'' and ``machine learning''. To this end, we propose to incorporate the meaning of the technologies into MF while extending the model capacity by passing the technology embedding through several linear layers. In the following, we describe in detail the layers of our architecture.%as illustrated in \autoref{}.

\sititle{Semantic embedding} We capture the technology meaning using BERT~\cite{sanh2019distilbert,devlin2018bert} as a feature extractor over the technology abstract:
\[\vs^{(0)}_i = f_{BERT}(a_i)\]
where $\vs_i$ is the semantic technology embedding and $a_i$ is the abstract of technology $t_i$. 

\sititle{MLP layers} The semantic technology embedding is passed through several MLP layers to further reduce the size of the embedding while increasing the model capacity. 

\begin{equation}\label{eq1}
  \begin{gathered}
	\vs^{(1)}_i = \mW_1 \vs^{(0)}_i + b_1\\
%	\vs^{(2)}_i = \sigma(\mW_2 \vs^{(1)}_i + b_2)\\
	\cdots \\
	\vs^{(k)}_i = \mW_k \vs^{(k-1)}_i + b_k
  \end{gathered}
\end{equation}
where $\mW_i, b_i, \sigma$ are the weight matrix, the bias and the non-linearity. 

\sititle{Combination layer} To obtain the final technology embedding $\vt_i$, we combine the semantic technology embedding $\vs^{(k)}$ with the raw technology embedding from MF $\ve_i$ by summing them: $\vt_i = \vs^{(k)} + \ve_i$. The summation allows us to save model's parameters in comparison with concatenation while it is also inspired by transformer architecture~\cite{vaswani2017attention,devlin2018bert} where positional encodings are added to the word embeddings. 

%We represent each company and technology by an embedding. The company and technology embeddings are learned based on the interaction matrix $\mM$. The technology embeddings, in addition, need to capture the connection between technologies as technologies are semantically connected. To this end, we propose to construct technology embeddings using phrase embedding. We use BERT phrase embedding to initialize the technology embedding $\vt$. The technology embeddings are forward through several layers to connect with the company embedding to predict the interaction matrix $\mM$.

%On the other hand, the company embeddings are randomly initialized and they are updated during the learning process. 

%In addition, in our setting, the technologies are semantically connected

\sstitle{Model learning} To learn the parameters, traditional MF uses the squared loss between the predicted and actual interaction score: $\Ls_{sqr}(\Theta) = \sum_{c_i,t_j \in \mM} (\emM_{ij} - \hat{\emM}_{ij})$.
However, such a method does not consider the unobserved entries directly. To this end, we follow the pairwise learning approach that aims to optimize the relative ranking between technologies. We use the margin hinge loss which is defined as follows:
\[
\Ls_{hinge}(\Theta, c_i, t_j, t_k) = max(0, m + \emM_{ij} - \hat{\emM}_{ik})
\]
where $c_i, t_j$ is an observed pair of company and technology while $t_k$ is a negative sample meaning $c_i, t_k$ is an unobserved entry of the interaction matrix $\mM$. 
%Traditionally, parameters of collaborative filtering model are learned based on pointwise loss function such as BPR~\cite{}. In our case, we use the hinge loss for its ability to capture the ...

\sstitle{Recommendation-based Retrieval} Then, the com-tech retrieval problem can be answered by ranking the technologies $\sT$ with respect to a company $c$ based on their interaction scores. 
More precisely, let $\mN$ to be the com-tech interaction matrix after the unobserved entries are estimated. 
The top-$k$ com-tech retrieval result for a company $c_i$ is a list of technologies ordered by their interaction scores $\mN$. 
%\todo{$R_i = (t_{\pi_1},t_{\pi_2},\cdots,t_{\pi_k})$} of technologies ordered by their interaction scores $\mN$ where $\pi$ is a permutation. 

\label{sec:retrieval}

\section{Empirical Evaluation}
\label{sec:eval}

\subsection{Experimental setup}

\sstitle{Datasets} We evaluate our model on the constructed dataset described in Section 3. As our dataset is crawl constantly which makes it difficult for evaluation, we fix the dataset used in the experiments to be the data collected before 01/04/2020. {This snapshot and the code are publicly available at \url{https://figshare.com/s/c014bb8565705e74dd1b}. %\thang{Should we publish the dataset right away or wait for the paper to be accepted?}

\sstitle{Metrics} 
To evaluate the results of com-com retrieval, we leverage the company categorisation from Crunchbase.
We measure the quality of com-com retrieval by the number of overlapping categories between the query company and the results. More precisely, let $\sC(c)$ denote the set of categories of company $c$. We define the retrieval accuracy for a company $c$ considering top-$k$ most relevant results as follows: 
\begin{equation}
P@k(c) = \frac{\sum_{i=\overline{1,k}} |\sC(c) \cup \sC(c_i)|}{k}
\label{equ:acc}
\end{equation}
This metric can be extended to a set of companies $C$ as $P@k(C) = \frac{\sum_{c \in C}P@k(c)}{|C|}$.

While it is ``straightforward'' to evaluate the search results for com-com, the evaluation for com-tech and tech-com is more challenging as there is no available ground truth. 
For tech-com search, we follow the approach of com-com retrieval where we label each technology by the Crunchbase categories. This is akin to consider each technology as a ``company''. 
The number of technologies to be labelled is usually small as we are interested in only important technologies. To this end, we have labeled 119 technologies which are considered important in the cybersecurity domain~\cite{STIB}. 
The retrieval accuracy $P@k(t)$ for a technology $t$ is defined similarly as in \autoref{equ:acc}. 
On the other hand, for com-tech retrieval, this approach is not practical as for each company, the list of retrieved technologies is very large. To this end, we opt for a qualitative evaluation.

\sstitle{Baselines} For technology classification, we construct a baseline using SVM on tf-idf featurization. 
More precisely, we construct a vector representing a Wikipedia category by combining the vector distances of a category to each of the Wikipedia's MTC and its TF-IDF weighted bag of words (BOW) representation. The weighted BOW representation of the given category is created from the stemmed text obtained by concatenating the abstracts of all Wikipedia articles directly connected to it. Mutual information based feature reduction than resulted in a vector of the length 1000.
These vectors are used as the input features for the classifier. 
For retrieval, we first compare with a tf-idf retrieval approach where the tf-idf values are also the relevancy of the technologies in com-tech retrieval. For com-com retrieval, it is an tf-idf weighted version of Jaccard similarity where each company is represented by its set of technologies. We also compare our recommendation-based retrieval approach with other recommendation models including GMF~\cite{he2017neural,rendle2009bpr}, MLP~\cite{he2017neural} and NCF~\cite{he2017neural}. {The above baselines are selected as to the best of our knowledge, there is no public implementation of technology classification and retrieval techniques. The above baselines represent the best starting points for these tasks.}

\sstitle{Environments} Our experiments ran on an Intel Xeon CPU E5-2620 v4 @ 2.10GHz server with a Titan V GPU with 12GB VRAM and 128GB RAM. Our model was implemented using Pytorch 1.7.1 and Spotlight as the recommendation framework and DistilBERT from HuggingFace as the language model. %The code is available at \todo{link}. 

\subsection{Effectiveness of technology classification}

\sstitle{Quality of technology classifiers} In this experiment, we analyze the correctness of our technology classifiers. We compare our proposed classifier using BERT with the baseline classifiers based on tf-idf and other BERT models. For this experiment, we compare these approaches on three metrics: accuracy, f1-score and AUC. We use k-fold cross validation with a 80-20 split.
%We expect our approach which uses modern language models is better than the baseline. 
\begin{table}[!th]
  \vspace{-0.5em}
\caption{Comparison of technology classifiers\label{tbl:class}}
\centering
\begin{tabular}{@{}lccc@{}}
\toprule
                          & F1-score                         & Accuracy                         & AUC                              \\ \midrule
tf-idf                    & 0.531 & 0.781 & 0.819 \\
DistilBERT w/o refine w/o finetune &   0.499 & 0.771 & 0.811                              \\
%\midrule
DistilBERT + refine + w/o finetune &   0.597 & 0.734 & 0.806                              \\
FDistilBERT + refine + w/ finetune (Ours)  & \textbf{0.639} & \textbf{0.799} & \textbf{0.857} \\ \bottomrule
\end{tabular}
  \vspace{-1.5em}
\end{table}

The results show in \autoref{tbl:class} confirms the benefit of fine-tuning and our refinement step. Our proposed approach outperforms the baselines on all metrics. The difference between using tf-idf as feature and BERT is 0.1, 0.02 and 0.04 for F1-score, accuracy and AUC respectively. This is expected as large language models trained on large text corpora are able to capture word meanings better. We also observe that fine-tuning improves accuracy in comparison with using BERT without fine-tuning.

%\sstitle{Effects of technology classifiers} In the next experiment, we aim to understand the effects of technology classifiers on the retrieval results. To this end, we vary the classifier quality by controlling the \todo{...}. For each classifier, we obtain a different set of technologies which we use as input for our retrieval model. The results in \autoref{} shows that there is an increase in tech-com retrieval quality as we improve the classifier quality. This is expected as better classifiers remove more noise (e.g non-technological entities), which means better company representation for better retrieval. 

\subsection{Effectiveness of technology retrieval}
In this experiment, we analyze our proposed recommendation-based retrieval model. We compare our model with a tf-idf based retrieval where each technology is associated with a tf-idf value while each company is represented by a tf-idf vector. We also compare our approach with several recommendation models including 1) Generalized MF~\cite{rendle2009bpr} which is a generalized version of MF, 2) MLP~\cite{he2017neural} which is a multi-layer recommendation model starting from random vectors and 3) NCF~\cite{he2017neural} or Neural Collaborative Filtering which is a recommendation model based on deep learning.

\begin{table*}[!th]
  \vspace{-0.5em}
\caption{Effects of retrieval model}
\label{tbl:retrieval}
\centering
%\resizebox{\textwidth}{!}{
\begin{tabular}{@{}lccccccccc@{}}
\toprule
        & \multicolumn{4}{l}{Company-company retrieval} & \multicolumn{4}{l}{Technology-company retrieval} \\ 
                \cmidrule(lr){2-5} \cmidrule(lr){6-9}
        & top-5      & top-10     & top-15     & top-20     & top-5       & top-10      & top-15      & top-20     \\
\midrule
MF     & 3.121    & 3.556    & 3.568    & 3.568    & 1.078     & 1.882     & 2.816     & 3.461    \\
MLP & 3.215    & 3.819    & 3.836    & 3.836    & 1.065     & 2.053     & 3.079     & 3.539    \\
NCF     & 2.905    & 3.103    & 3.105    & 3.105       & 1.118     & 2.211     & 2.947     & 3.5    \\
\midrule
BERT    & 4.155    & 5.986    & 6.636    & 6.722    & 1.986     & 3.211     & 3.882     & 5.066    \\
MF+BERT & \textbf{4.458} & \textbf{7.004} & \textbf{8.447} & \textbf{8.845} & \textbf{2.197} & \textbf{3.684} & \textbf{5.105} & \textbf{6.289}   \\ \bottomrule
\end{tabular}
%}
  \vspace{-1.5em}
\end{table*}

The experimental results shown in \autoref{tbl:retrieval} show that our proposed model based on BERT embeddings as initial technology embeddings are better than the baselines. The difference between our worst model and the best baseline is 0.6 at top-5 for com-com retrieval and 1.8 at top-5 for tech-com retrieval. 
This can be explained by the fact that our models can capture the meaning of the technologies while the baselines consider the technologies to be independent. Among our proposed models, adding the raw technology embedding from MF with the BERT technology embedding is better than using the BERT embedding alone. We can attribute this to the larger capacity of our model which helps in capture the interaction matrix better.

\subsection{Parameter sensitivity}

\begin{figure}[!hbt]
\centering
  \vspace{-0.5em}
  \includegraphics[width=1.0\linewidth]{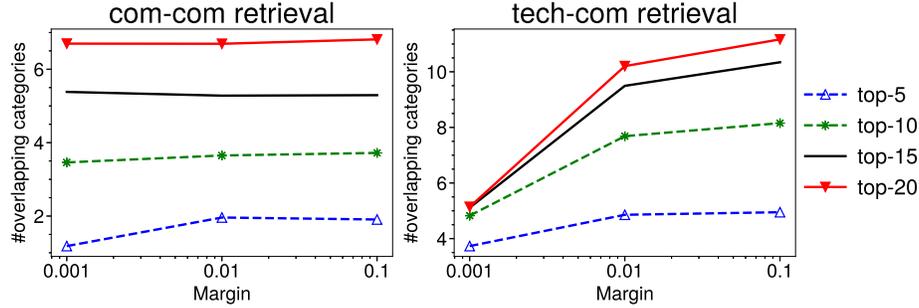}
    \vspace{-1.5em}
  \caption{Effects of margin}
  \label{fig:margin}
  \vspace{-1.5em}
\end{figure}

\sstitle{Effects of margin} We vary the margin of the hinge loss from 0.001 to 0.1 to analyze its effects on the retrieval results. The experimental results are shown in \autoref{fig:margin}. We observe that the number of overlapping categories tends to increase with the margin. For instance, the number of overlapping categories for top-5 com-com retrieval is 3.73 when the margin is 0.001 but it increases to 4.95 when the margin is 0.1. 
This is expected as with the larger margin, our model tends to generalize as it aims to capture common technologies between the companies. 
We observe this phenomenon clearly from \autoref{tbl:comtech} that we obtain more specific technologies with smaller margin. 
With larger margins, generic technologies that are shared among different companies are more representative than more specific ones. This experiment confirms our ability to control the specificity of the retrieval results by changing the margin of the hinge loss. 
%For tech-com retrieval, an interesting observation is that the results do not improve as we increase the margin from 1 to 2. 

\begin{figure}[!hbt]
\centering
  \vspace{-0.5em}
  \includegraphics[width=1.0\linewidth]{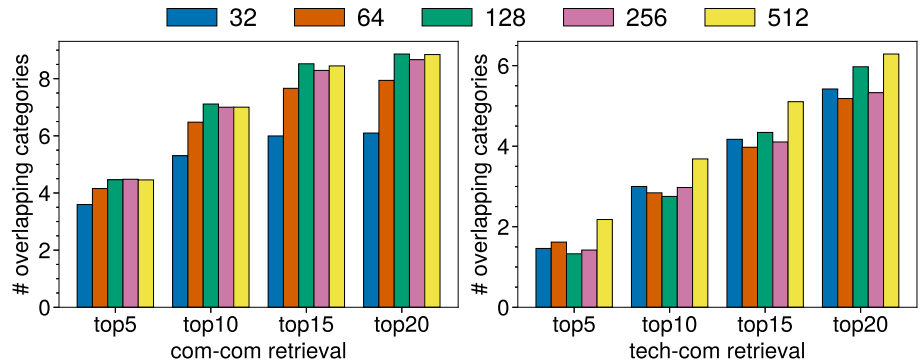}
    \vspace{-1.5em}  
  \caption{Effects of embedding size}
  \label{fig:emb}
  \vspace{-2.5em}  
\end{figure}

%There is a slight decrease for com-com retrieval at top-5,10 and 20 when the margin increases from 0.1 to 1.

\sstitle{Embedding size} In this experiment, we analyse the effects of the embedding size on the retrieval results. We vary the embedding size from 32 to 512. Results in \autoref{fig:emb} shows that as the embedding size increases, we can retrieve companies better for both tech-com and com-com retrieval tasks. This is expected as increasing embedding size also improves the model capacity. However, there is a trade-off in increasing embedding size as it incurs longer training time as shown in \autoref{fig:time}. The difference in training time between embedding size of 32 and 512 is 3 times. However, even with the largest embedding size, the training time per epoch is still very fast - only around 1.5 second.

\begin{figure}[h]
  \vspace{-0.5em}
	\begin{minipage}{.47\linewidth}
	\centering
	\includegraphics[width=0.9\linewidth]{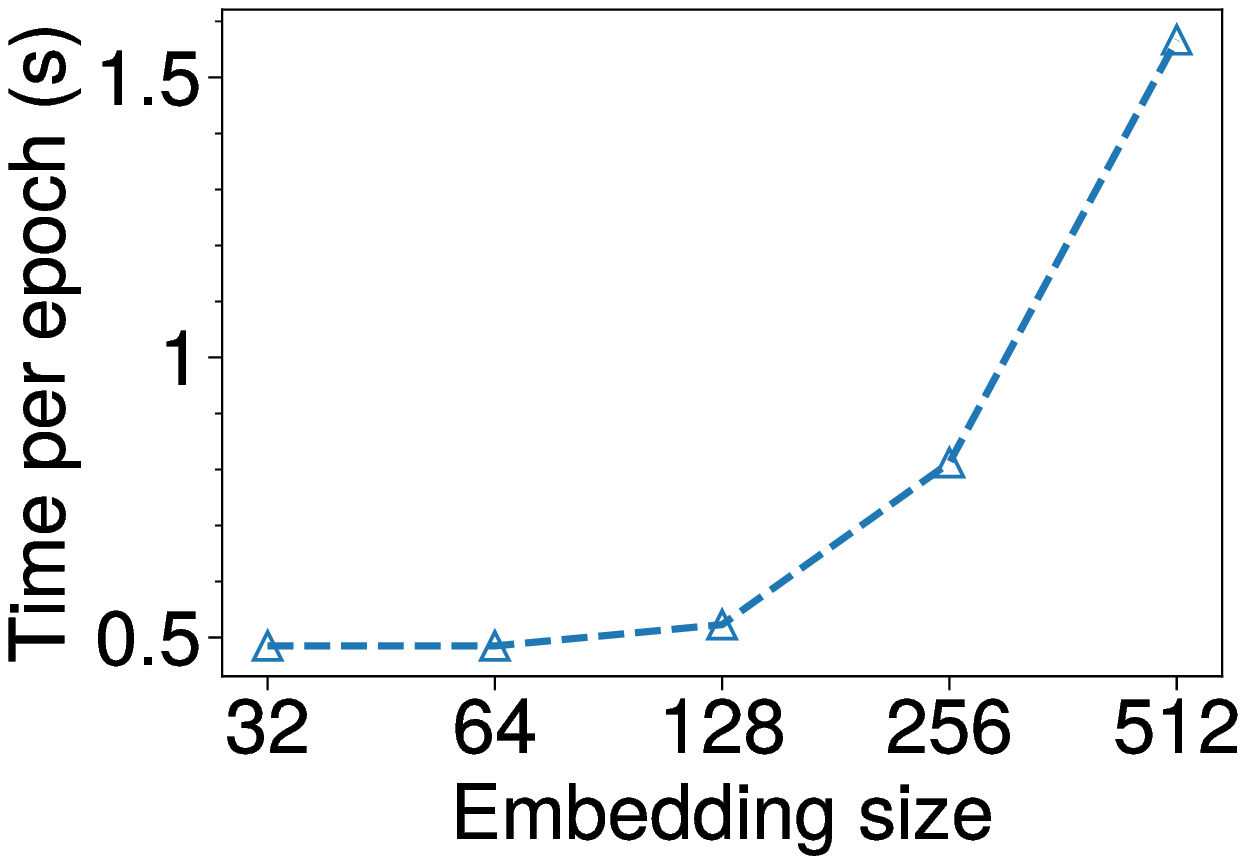}
	\vspace{-0.9em}
	\caption{Training time per epoch.}
	\label{fig:time}
\end{minipage}
	\quad
	\begin{minipage}{.47\linewidth}
		\centering

\begin{tabular}{@{}lcccc@{}}
\toprule
      & \multicolumn{2}{l}{com-com retrieval} & \multicolumn{2}{l}{tech-com retrieval} \\                 \cmidrule(lr){2-3} \cmidrule(lr){4-5}
      & tf-idf                   & Ours                  & tf-idf                      & Ours                    \\ \midrule
top-5  & 1.2916         &  \textbf{4.458}        & 2.0161             & \textbf{2.197}            \\
top-10 & 1.3022         & \textbf{7.004}        &  3.4838             &  \textbf{3.648}            \\
top-15 & 1.3022         &  \textbf{8.447}        & 4.3710             &  \textbf{5.105}            \\
top-20 & 1.3022         & \textbf{8.845}        &  4.7419             &  \textbf{6.289}            \\ \bottomrule
\end{tabular}
\vspace{0.5em}
\captionof{table}{End-to-end evaluation}
\label{tbl:end2end}
	\end{minipage}
	  \vspace{-2.em}
\end{figure}

\subsection{End-to-end comparison} Having evaluated the individual components of our solution, we turn to its end-to-end performance in comparison with the baseline. \autoref{tbl:end2end} compares the performance of our approach with a tf-idf based retrieval approach which uses tf-idf as feature for technology classifier and technology retrieval. Our approach leads to significantly better retrieval results in both retrieval tasks. Our model is nearly 4 times better than the baseline in the com-com retrieval at top-5 while the difference is 0.18 for tech-com retrieval. First, this can be attributed to our approach's better technology classifier by using language model. Second, our recommendation retrieval model considers both the companies, technologies and their relationships as the same time, while the technologies and companies in the tf-idf model are handled independently. This enables our model to leverage the similarity of companies to support technology retrieval and vice versa.

%
%% Please add the following required packages to your document preamble:
%% \usepackage{booktabs}
%\begin{table}[!th]
%\centering
%\caption{End-to-end evaluation}
%\label{tbl:end2end}
%\begin{tabular}{@{}lcccc@{}}
%\toprule
%      & \multicolumn{2}{l}{com-com retrieval} & \multicolumn{2}{l}{tech-com retrieval} \\                 \cmidrule(lr){2-3} \cmidrule(lr){4-5}
%      & tf-idf                   & Ours                  & tf-idf                      & Ours                    \\ \midrule
%top-5  & 1.2916         &  \textbf{4.458}        & 2.0161             & \textbf{2.197}            \\
%top-10 & 1.3022         & \textbf{7.004}        &  3.4838             &  \textbf{3.648}            \\
%top-15 & 1.3022         &  \textbf{8.447}        & 4.3710             &  \textbf{5.105}            \\
%top-20 & 1.3022         & \textbf{8.845}        &  4.7419             &  \textbf{6.289}            \\ \bottomrule
%\end{tabular}
%\end{table}

\begin{table}[!th]
  \vspace{-0.5em}
\caption{Com-tech retrieval\label{tbl:comtech}}
\scalebox{0.8}{
\begin{tabular}{@{}llllll@{}}
\toprule
\multicolumn{3}{c}{Acronis AG}                                   & \multicolumn{3}{c}{InterHype SARL}                               \\ \cmidrule(lr){1-3}\cmidrule(lr){4-6}
Ours-0.01                & Ours-0.1          & tf-idf            & Ours-0.01                  & Ours-0.1               & tf-idf     \\
\midrule
CyberTruck               & Encryption        & Cloud computing   & Computer security          & Cloud computing        & Nous       \\
Virtual machine          & Communication     & Backup            & Automatic train protection & Computer science       & Sand       \\
Cloud storage            & Virtualization    & Disaster recovery & SMS                        & Computer security      & Antiseptic \\
Off-site data protection & Internet          & Web server        & Off-site data protection   & Digital transformation & Habitat    \\
Encryption               & Personal firewall & Ransomware        & Backup                     & Information security   & Glass      \\ \bottomrule
\end{tabular}
}
  \vspace{-2.5em}
\end{table}

\subsection{Qualitative analysis}
In this experiment, we analyze the retrieval results qualitatively. \autoref{tbl:comtech} shows the com-tech retrieval results where we search for cyber security companies. For com-tech retrieval, as discussed above, we are able to control the specificity of the results by changing the margin. In addition, our proposed model is able to return less noise in comparison with tf-idf one as tf-idf model may return non-technological terms due to the quality of its classifier. This phenomenon can be seen for instance in the search for InterHype Sarl which is a cyber security company. For com-com and tech-com retrieval, due to space constraint, we do not include them. However, we observe that our model can return companies that are in the same domains as the queried company or technology. This is in line with the quantitative result observed in previous experiments.

%For instance, for InterHype com-tech search, tf-idf model returns non-technological terms as its classifier 

\section{Conclusion and Future works}
In this paper, we propose an end-to-end framework to first extract and classify technological mentions from company corpuses and then, retrieve related technologies and companies. Our technology classifier is based on DistilBERT model with finetuning and refinement to achieve better accuracy while our recommendation-based retrieval model enables more relevant results. In the future works, we aim to allow users to reformulate queries to better capture users intentions. Another possible research direction is to combine technology classification and entity extraction for more accurate results.

%\dimitri{Summary and hands-on utility (for the monitoring of the technological landscape: I suggest Alain can write something here} \alain{inputs}

%Limitations: 1) current approach does not allow users to reformulate queries. 2) tech classification and entity extraction are done separately which makes tech classification lose contexts. 

%Some future works: 1) add query formulation component to better capture user intention 2) combine tech classification in entity extraction step to extract technologies directly with better accuracy. 
\label{sec:conclusion}

\bibliographystyle{splncs04}
%\printbibliography
\bibliography{biblio.bib}

\begin{thebibliography}{10}
\providecommand{\url}[1]{\texttt{#1}}
\providecommand{\urlprefix}{URL }
\providecommand{\doi}[1]{https://doi.org/#1}

\bibitem{STIB}
Security-relevant technology and industry base.
  \url{https://www.ar.admin.ch/en/beschaffung/ruestungspolitik-des-bundesrates/offset.html},
  accessed: 2021-06-16

\bibitem{aggarwal2018information}
Aggarwal, C.C.: Information retrieval and search engines. In: Machine Learning
  for Text, pp. 259--304. Springer (2018)

\bibitem{aharonson2016mapping}
Aharonson, B.S., Schilling, M.A.: Mapping the technological landscape:
  Measuring technology distance, technological footprints, and technology
  evolution. Research Policy  \textbf{45}(1),  81--96 (2016)

\bibitem{Arai2020}
Arai, K.: Extraction of keywords for retrieval from paper documents and
  drawings based on the method of determining the importance of knowledge by
  the analytic hierarchy process: Ahp. International Journal of Advanced
  Computer Science and Applications  \textbf{11}(10) (2020)

\bibitem{10.1561/1500000024}
Balog, K., Fang, Y., de~Rijke, M., Serdyukov, P., Si, L.: Expertise retrieval
  \textbf{6}(2-3),  127--256 (Feb 2012)

\bibitem{belkin1992information}
Belkin, N.J., Croft, W.B.: Information filtering and information retrieval: Two
  sides of the same coin? Communications of the ACM  \textbf{35}(12),  29--38
  (1992)

\bibitem{chen2019disruptive}
Chen, X., Han, T.: Disruptive technology forecasting based on gartner hype
  cycle. In: TEMSCON. pp.~1--6. IEEE (2019)

\bibitem{10.1145/2506182.2506198}
Daiber, J., Jakob, M., Hokamp, C., Mendes, P.N.: Improving efficiency and
  accuracy in multilingual entity extraction. In: ICSS. pp. 121--124. New York,
  NY, USA (2013)

\bibitem{daim2016anticipating}
Daim, T.U., Chiavetta, D., Porter, A.L., Saritas, O.: Anticipating future
  innovation pathways through large data analysis. Springer (2016)

\bibitem{demartini2009vector}
Demartini, G., Gaugaz, J., Nejdl, W.: A vector space model for ranking entities
  and its application to expert search. In: ECIR. pp. 189--201. Springer (2009)

\bibitem{devlin2018bert}
Devlin, J., Chang, M.W., Lee, K., Toutanova, K.: Bert: Pre-training of deep
  bidirectional transformers for language understanding. arXiv preprint
  arXiv:1810.04805  (2018)

\bibitem{dotsika2017identifying}
Dotsika, F., Watkins, A.: Identifying potentially disruptive trends by means of
  keyword network analysis. Technological Forecasting and Social Change
  \textbf{119},  114--127 (2017)

\bibitem{durak2018flight}
Durak, U.: Flight 4.0: The changing technology landscape of aeronautics. In:
  Advances in Aeronautical Informatics, pp. 3--13. Springer (2018)

\bibitem{he2017neural}
He, X., Liao, L., Zhang, H., Nie, L., Hu, X., Chua, T.S.: Neural collaborative
  filtering. In: WWW. pp. 173--182 (2017)

\bibitem{heggo2021data}
Heggo, I.A., Abdelbaki, N.: Data-driven information filtering framework for
  dynamically hybrid job recommendation. In: Intelligent Systems in Big Data,
  Semantic Web and Machine Learning, pp. 23--49. Springer (2021)

\bibitem{hersh2014information}
Hersh, W.R.: Information retrieval and digital libraries. In: Biomedical
  Informatics, pp. 613--641. Springer (2014)

\bibitem{hossari2019test}
Hossari, M., Dev, S., Kelleher, J.D.: Test: A terminology extraction system for
  technology related terms. In: Proceedings of the 2019 11th International
  Conference on Computer and Automation Engineering. pp. 78--81 (2019)

\bibitem{kaur2021comparative}
Kaur, P., Pannu, H.S., Malhi, A.K.: Comparative analysis on cross-modal
  information retrieval: a review. Computer Science Review  \textbf{39},
  100336 (2021)

\bibitem{kim2016generating}
Kim, M., Park, Y., Yoon, J.: Generating patent development maps for technology
  monitoring using semantic patent-topic analysis. Computers \& Industrial
  Engineering  \textbf{98},  289--299 (2016)

\bibitem{loveridge2016fta}
Loveridge, D., Cagnin, C.: Fta as due diligence for an era of accelerated
  interdiction by an algorithm-big data duo. In: Anticipating Future Innovation
  Pathways Through Large Data Analysis, pp. 3--23. Springer (2016)

\bibitem{madnick2008semantic}
Madnick, S., Woon, W.L.: Semantic distances for technology landscape
  visualization  (2008)

\bibitem{mahmood2013security}
Mahmood, T., Afzal, U.: Security analytics: Big data analytics for
  cybersecurity: A review of trends, techniques and tools. In: 2013 2nd
  national conference on Information assurance (ncia). pp. 129--134. IEEE
  (2013)

\bibitem{mendes2011dbpedia}
Mendes, P.N., Jakob, M., et~al.: Dbpedia spotlight: shedding light on the web
  of documents. In: Proceedings of the 7th international conference on semantic
  systems. pp.~1--8 (2011)

\bibitem{mikheev2018ontology}
Mikheev, A.: Ontology-based data access for energy technology forecasting. In:
  IWCI 2018. pp. 147--151. Atlantis Press (2018)

\bibitem{munir2018use}
Munir, K., Anjum, M.S.: The use of ontologies for effective knowledge modelling
  and information retrieval. Applied Computing and Informatics  \textbf{14}(2),
   116--126 (2018)

\bibitem{rao2019multi}
Rao, J., Yang, W., Zhang, Y., Ture, F., Lin, J.: Multi-perspective relevance
  matching with hierarchical convnets for social media search. In: Proceedings
  of the AAAI Conference on Artificial Intelligence. vol.~33, pp. 232--240
  (2019)

\bibitem{rendle2009bpr}
Rendle, S., Freudenthaler, C., Gantner, Z., Schmidt-Thieme, L.: Bpr: Bayesian
  personalized ranking from implicit feedback. In: UAI. pp. 452--461 (2009)

\bibitem{rikhardsson2018business}
Rikhardsson, P., Yigitbasioglu, O.: Business intelligence \& analytics in
  management accounting research: Status and future focus. International
  Journal of Accounting Information Systems  \textbf{29},  37--58 (2018)

\bibitem{roetzel2019information}
Roetzel, P.G.: Information overload in the information age: a review of the
  literature from business administration, business psychology, and related
  disciplines with a bibliometric approach and framework development. Business
  research  \textbf{12}(2),  479--522 (2019)

\bibitem{ruder2016overview}
Ruder, S.: An overview of gradient descent optimization algorithms. arXiv
  preprint arXiv:1609.04747  (2016)

\bibitem{sanh2019distilbert}
Sanh, V., Debut, L., Chaumond, J., Wolf, T.: Distilbert, a distilled version of
  bert: smaller, faster, cheaper and lighter. arXiv preprint arXiv:1910.01108
  (2019)

\bibitem{shalaby2019patent}
Shalaby, W., Zadrozny, W.: Patent retrieval: a literature review. Knowledge and
  Information Systems pp. 1--30 (2019)

\bibitem{shalf_future_2020}
Shalf, J.: The future of computing beyond {Moore}'s {Law}. Philosophical
  Transactions of the Royal Society A: Mathematical, Physical and Engineering
  Sciences  \textbf{378}(2166) (2020)

\bibitem{sitarz2012application}
Sitarz, R., Kraslawski, A.: Application of semantic and lexical analysis to
  technology forecasting by trend analysis-thematic clusters in separation
  processes. In: Computer aided chemical engineering, vol.~30, pp. 437--441.
  Elsevier (2012)

\bibitem{tang2008applying}
Tang, F., Liu, Y.: Applying semantic web into technology forecasting in
  enterprises. In: International Conference on Service Operations and
  Logistics, and Informatics. vol.~1, pp. 135--138. IEEE (2008)

\bibitem{vaswani2017attention}
Vaswani, A., Shazeer, N., Parmar, N., Uszkoreit, J., Jones, L., Gomez, A.N.,
  Kaiser, {\L}., Polosukhin, I.: Attention is all you need. NeurIPS
  \textbf{30},  5998--6008 (2017)

\bibitem{wang2017leveraging}
Wang, Y., Rastegar-Mojarad, M., Komandur-Elayavilli, R., Liu, H.: Leveraging
  word embeddings and medical entity extraction for biomedical dataset
  retrieval using unstructured texts. Database  \textbf{2017} (2017)

\end{thebibliography}
%bibitem{ref_article1}
%Author, F.: Article title. Journal \textbf{2}(5), 99--110 %2016)

%uthor, F., Author, S.: Title of a proceedings paper. In: Editor,
%F., Editor, S. (eds.) CONFERENCE 2016, LNCS, vol. 9999, pp. 1--13.
%Springer, Heidelberg (2016). \doi{10.10007/1234567890}

%Author, F., Author, S., Author, T.: Book title. 2nd edn. Publisher,

%Author, A.-B.: Contribution title. In: 9th International Proceedings
%on Proceedings, pp. 1--2. Publisher, Location (2010)

%LNCS Homepage, \url{http://www.springer.com/lncs}. Last accessed 4
%\end{thebibliography}
\end{document}